\documentclass[preprint,12pt]{elsarticle}

\usepackage{amsmath,amssymb,amsfonts,amsthm}
\usepackage[english]{babel}
\usepackage{url}
\usepackage{bm}
\usepackage[latin1]{inputenc}
\usepackage{graphicx}
\usepackage{hyperref}
\usepackage{multirow, array}
\usepackage{caption}
\usepackage{subcaption}
\usepackage{listings}

\usepackage{dcolumn}

\usepackage[perpage,bottom]{footmisc}

\usepackage[usenames,dvipsnames]{xcolor}

\lstdefinelanguage{Ini}
{
    basicstyle=\ttfamily\small,
    columns=fullflexible,
    morecomment=[s][\color{Orchid}\bfseries]{[}{]},
    morecomment=[l]{\#},
    morecomment=[l]{;},
    commentstyle=\color{gray}\ttfamily,
    morekeywords={},
    otherkeywords={=,:}
}

\begin{document}

\begin{frontmatter}

\title{Beta Spectrum Generator: High precision allowed $\beta$ spectrum shapes}

\author[a]{L. Hayen\corref{author}}
\author[a]{N. Severijns}

\cortext[author] {Corresponding author.\\\textit{E-mail address:} leendert.hayen@kuleuven.be}
\address[a]{Instituut voor Kern- en Stralingsfysica, KU Leuven, Celestijnenlaan 200D, B-3001 Leuven, Belgium}

\date{\today}
\begin{abstract}
Several searches for Beyond Standard Model physics rely on an accurate and highly precise theoretical description of the allowed $\beta$ spectrum. Following recent theoretical advances, a C++ implementation of an analytical description of the allowed beta spectrum shape was constructed. It implements all known corrections required to give a theoretical description accurate to a few parts in $10^4$. The remaining nuclear structure-sensitive input can optionally be calculated in an extreme single-particle approximation with a variety of nuclear potentials, or obtained through an interface with more state-of-the-art computations. Due to its relevance in modern neutrino physics, the corresponding (anti)neutrino spectra are readily available with appropriate radiative corrections. In the interest of user-friendliness, a graphical interface was developed in Python with a coupling to a variety of nuclear databases. We present several test cases and illustrate potential usage of the code. Our work can be used as the foundation for current and future high-precision experiments related to the beta decay process.\\
Source code: \url{https://github.com/leenderthayen/BSG}\\
Documentation: \url{http://bsg.readthedocs.io}
\end{abstract}

\end{frontmatter}

{\bf Program summary} \\

\begin{small}
\noindent
{\em Program Title: } BSG                                          \\
{\em Licensing provisions: } MIT                                    \\
{\em Programming language: } C++ and Python                                   \\
%
{\em Nature of problem:}
The theoretical allowed $\beta$ spectrum contains a large variety of corrections from different areas of physics, each of which is important in certain energy ranges. A high precision description is required for new physics searches throughout the entire nuclear chart. \\
{\em Solution method:}
We implement the analytical corrections described in recent theoretical work. Nuclear matrix elements in allowed Gamow-Teller $\beta$ decay are calculated in a spherical harmonic oscillator basis. Wave functions can be calculated in an extreme single-particle approximation using different nuclear potentials, or provided by the user as the output from more sophisticated routines. Corresponding neutrino spectra are calculated with appropriate radiative corrections. A graphical user interface written in Python additionally provides connections to a variety of nuclear databases.\\
{\em CPC Library subprograms used:}

{\em Catalogue number:} ABOV

{\em Title:} NILSSON ORBITS

{\em Ref. in CPC:} 6 (1973) 30
   
\end{small}

\section{Introduction}
Precision studies of the $\beta$ decay process have shaped the evolution of the Standard Model, and continue to be at the forefront of new physics searches. It offers a wide variety of sensitive, experimental observables, ranging from correlations to the actual spectrum shape \cite{Severijns2014}. For example, the most precise determination of $V_{ud}$, the up-down quark element of the Cabibbo-Kobayashi-Maskawa matrix, is determined from the integrated form of the $\beta$ spectrum \cite{Hardy2015} while search for exotic physics is both competitive and complementary to LHC searches \cite{Bhattacharya2012, Naviliat-Cuncic2013}. As experiments begin exploring ever higher-precision regimes \cite{Sternberg2015, Fenker2018, Vos2015}, the need for an equally improved theoretical formalism taking into account all the intricacies of a complex many-body system is clear. This is reinforced by outstanding anomalies in the reactor neutrino sector and the cardinal role played by the $\beta$ spectrum \cite{Hayes2016}. 

In the classic $\beta$ decay experiments one is sensitive to exotic currents in the weak interaction, typically of the scalar or tensor sort \cite{Holstein2014a}. Following the results of Ref. \cite{Jackson1957}, the presence of these currents modify the spectrum through the so-called Fierz term in the following way
\begin{equation}
N(W)dW \propto 1 + \frac{\gamma m_e}{W}b_\text{Fierz},
\label{eq:spectrum_fierz}
\end{equation}
where $\gamma = \sqrt{1 - (\alpha Z)^2}$ with $\alpha$ the fine-structure constant and $Z$ the proton number, $m_e$ the electron mass and $W$ the electron total energy. The Fierz contribution is given by
\begin{equation}
b_\text{Fierz} = \pm \frac{1}{1+\rho^2}\left[ \text{Re}\left( \frac{C_S+C_S'}{C_V}\right) + \rho^2 \text{Re} \left(\frac{C_T+C_T'}{C_A} \right) \right],
\label{eq:fierz}
\end{equation}
where $C_i^{(\prime)}$ are the coupling constants of the usual Lee-Yang Hamiltonian \cite{Lee1956} and $\rho = \frac{C_AM_{GT}}{C_VM_F}$ is the Fermi / Gamow-Teller mixing ratio. In the Standard Model Eq. (\ref{eq:fierz}) is clearly zero, and so any observation of such a shape dependence would indicate new physics. It is then of the utmost importance to properly control all energy-dependent terms to a higher precision than what is required experimentally. 

The largest uncertainty comes from the remaining nuclear structure-sensitive input into the spectrum shape. The most substantial of these come from the currents induced into the weak Hamiltonian due to QCD intrusions. This effect can to first order be reduced to the influence of the so-called weak magnetism term, such that the spectrum is modified roughly in the following way

\begin{equation}
N(W)dW \propto1 + \frac{\gamma m_e}{W} b_{\text{Fierz}} \pm \frac{4}{3} \frac{W}{M} b_{\text{wm}},
\label{eq:spectral_mod_fierz_wm}
\end{equation}
where $M$ is the nuclear mass, and upper (lower) signs represent $\beta^-$ ($\beta^+$) decay. Typical values for $b_\text{wm}$ result in slopes of 0.5$\%\,$MeV$^{-1}$. To allow meaningful extraction of new physics results at the $10^{-3}$ level, its value then needs to be known with a better than $20\%$ accuracy. In Eq. (\ref{eq:spectral_mod_fierz_wm}) all other corrections have been omitted for the sake of clarity, even though all of these collectively should be known to at least the same level.

Recently, work performed by Hayen \emph{et al.} \cite{Hayen2018} expounded the current formalism and introduced several additional corrections to bring the analytical $\beta$ spectrum shape to a relative precision of a few parts in $10^4$. Here, a particular focus was placed upon a proper description of atomic effects, such as the screening and exchange processes. Nuclear structure corrections were thoroughly investigated, and its evaluation using single-particle wave functions was carefully discussed. 
This work describes the implementation of this formalism into a user-friendly software package. This comprises both the beta spectrum shape as well as a library dedicated to calculating nuclear matrix elements. We specifically allow incorporation of advanced many-body method results such as the nuclear shell model, making the latter a versatile package.

The paper is organised as follows. In Sec. \ref{beta-full} we discuss the formulation of the $\beta$ spectrum shape through a brief summary of the physical reasoning behind the main terms as well as the different correction terms, including an estimate of their magnitude. Section \ref{sec:nuclear_structure} discusses the structure of the nuclear form factors and the required elements of its evaluation. 
Following this, Sec. \ref{sec:BSG_options} discusses the different options in the spectrum generation, followed by those of the matrix element evaluation in Sec. \ref{sec:NME_options}. Section \ref{sec:usage_examples} discusses some test case results in both libraries. Finally, we provide some examples of the potential of the results presented here in the field of low energy particle physics, specifically in the search for new physics.

\section{Allowed beta spectrum shape}
\label{beta-full}
\subsection{Overview}
Nuclear $\beta$ decay is a non-trivial ordeal where all three forces in the Standard Model come together in a many-body system. It is modified both by QCD inside the nucleus, as well as through atomic physics which dictates the electromagnetic potential the $\beta$ particle is created in. The analytical description consists of a significant amount of terms, all of which are represented as individual multiplicative factors relative to the simple phase space shape. For an allowed decay, it is written as \cite{Hayen2018}
\begin{align}
N(W)dW & = \frac{G_V^2 V_{ud}^2}{2\pi^3} ~ F_{0}(Z, W) ~ L_0(Z, W) ~ U(Z, W) ~ D_\text{FS}(Z, W, \beta_2) ~ R_N(W, W_0, M)  \nonumber \\
 & ~~~~ \times ~ Q(Z, W,M) ~ R(W, W_0) ~ S(Z, W) ~ X(Z, W) ~ r(Z, W) \nonumber \\
 & ~~~~ \times ~ C(Z, W) ~ D_C(Z, W, \beta_2) ~ p W (W_0 - W)^2 ~ dW  \nonumber \\
 & \equiv \frac{G_V^2 V_{ud}^2}{2\pi^3} ~ K(Z, W, W_0, M) ~ N(Z, W, \beta_2) ~ p W (W_0 - W)^2 ~ dW .
\label{eq:full_expression}
\end{align}
\noindent Here, $Z$ is the proton number of the daughter nucleus, $W_0$ is the total $\beta$ particle energy at the spectrum endpoint, $p = \sqrt{W^2 - 1}$ the $\beta$ particle momentum and $G_V$ the vector coupling strength in nuclei\footnote{All quantities are in units natural to $\beta$ decay, i.e. $\hbar = c = m_e = 1$.}. Further, $V_{ud}$ is the $up$-$down$ quark matrix element of the CKM quark-mixing matrix as before. To clearly differentiate the origin of the effects, electromagnetic and kinematic corrections are combined into the factor $K$, while the nuclear structure dependent part is written as $N(Z, W, \beta_2)$. Here $\beta_2$ represents the quadrupole deformation of the nucleus. More generally, we consider deformations such that the nuclear surface is given by an ellipsoid
\begin{equation}
R(\theta, \phi) = R_0\left(1  + \sum_{n > 0} \beta_{2n} Y_{2n}^0(\theta, \phi)\right),
\label{eq:nuclear_surface_deformation}
\end{equation}
with $R_0$ the nuclear radius and $Y_{L}^M$ the usual spherical harmonics.

\subsection{Participating corrections}
\label{sec:participating_corrections}
The factors in Eq. (\ref{eq:full_expression}) originate from different areas of physics. As such it is useful to present a brief overview of the physical meaning behind each of these corrections in order to better comprehend their influence. Table \ref{table:overview_BSG} shows all effects included in our description of the $\beta$ spectrum shape with corresponding references to the factors used. When estimates of the magnitude of different corrections are given, these represent the relative change in the phase space integral after inclusion of the effect, unless mentioned otherwise. Here we present a summary of all factors included (more details are given in Ref. \cite{Hayen2018}):
\begin{table}[!ht]
\centering
\caption{Overview of the features present in the $\beta$ spectrum shape (Eq. (\ref{eq:full_expression})), and the effects incorporated into the Beta Spectrum Generator Code presented here. The magnitudes are listed as the maximal typical deviation for medium $Z$ nuclei with a few MeV endpoint energy. Some of these corrections fall off very quickly (e.g. the exchange correction, $X$) but can be sizeable in a small energy region. Varying $Z$ or $W_0$ can obviously allow for some migration within categories for several correction terms.}
{\renewcommand{\arraystretch}{1.2}
\begin{tabular}{c l l r}
\hline \hline
Item & Effect & Formula & Magnitude \\
\hline
1 & Phase space factor &  $pW(W_0-W)^2$ & \multirow{2}{*}{Unity or larger} \\
2 & Traditional Fermi function &  $F_0$ &\\
\hline
3 & Finite size of the nucleus &  $L_0$ & \multirow{5}{*}{$10^{-1}$-$10^{-2}$} \\
4 & Radiative corrections &  $R$ & \\
5 & Shape factor & $C$ & \\
6 & Atomic exchange &  $X$ & \\
7 & Atomic mismatch & $r$ & \\
\hline
8 & Atomic screening &  $S$ & \multirow{10}{*}{$10^{-3}$-$10^{-4}$}\\
9 & Shake-up & See item 7 & \\
10 & Shake-off & See item 7 & \\
11 & Isovector correction & $C_I$ & \\
12 & Recoil Coulomb correction & $Q$ & \\
13 & Diffuse nuclear surface & $U$ & \\
14 & Nuclear deformation & $D_\text{FS}$ \& $D_C$ & \\
15 & Recoiling nucleus &  $R_N$ & \\
16 & Molecular screening & $\Delta S_{\text{Mol}}$ \\
17 & Molecular exchange & Case by case \\
\hline
18 & Bound state $\beta$ decay & $\Gamma_b/\Gamma_c$ & \multirow{2}{*}{Smaller than $1\cdot 10^{-4}$} \\
19 & Neutrino mass & Negligible & \\
\hline \hline
\end{tabular}
}
\label{table:overview_BSG}
\begin{flushleft}
\end{flushleft}
\end{table}

\begin{itemize}

\item \textit{$F_0(Z, W)$: } The Fermi function takes into account the distortion of the electron radial wave function by the nuclear charge, i.e. the Coulomb interaction between the $\beta$ particle and the daughter nucleus. This is traditionally defined at the origin, where one takes a ratio of the electron density for both Coulomb-distorted and free wave functions. The force is attractive for $\beta^-$ decay, such that the electron density increases and subsequently increases the decay rate, and vice versa for $\beta^+$ decay. The calculation of the traditional Fermi function involves representing the nucleus as a point particle with charge $Z$, as it is the only available analytical solution of the Dirac equation in a Coulomb field. As a consequence the electron radial wave functions diverge slightly at the origin and are instead evaluated at the nuclear radius\footnote{In the case of a uniformly charged sphere we can write $R = \sqrt{\frac{5}{3}}\langle r^2 \rangle_\text{exp}^{1/2}$, where $\langle r^2 \rangle_\text{exp}^{1/2}$ is the experimentally measured rms charge radius.}, $R$. As with all electromagnetic corrections discussed here, the effect is strongest for the lowest $\beta$ energies. As the Compton wavelength increases, the sampling of the surrounding electric field increases and so provides a larger spectral distortion. The Fermi function has by far the largest influence on the spectral shape.

\item \textit{$L_0(Z, W)$: }
When considering instead of a point nucleus one of finite size, the electron and positron wave functions become finite at the origin. Any other description but a point charge is, however, not analytically solvable, and so a numerical correction factor, $L_0(Z, W)$, is introduced to be used in combination with the analytical Fermi function. This corresponds to using a uniformly charged sphere as a nuclear charge distribution.

\item \textit{$U(Z, W)$: }
Even though the inclusion of $L_0$ is a step in the right direction, approximating the charge distribution as a simple uniformly charged sphere is too crude an approximation. In reality the nuclear charge distribution is smeared out over a certain distance.
This in turn introduces an additional, smaller, correction term from replacing the uniform spherical charge distribution that defines $L_0(Z, W)$ with a more realistic one of the same $\langle r^2 \rangle^{1/2}$, such as, e.g., the Fermi distribution.

\item \textit{$R_N(W, W_0, M)$: }
Usually, approximating the nucleus as infinitely massive is sufficiently precise. For lighter nuclei, however, this is no longer warranted relative to the required precision. The change from a 2-body to 3-body process introduces an additional kinematic correction. As it is governed by conservation of momentum, it is different for Fermi and Gamow-Teller decays. All corrections described that are sensitive to the type of decay can simply be calculated by weighting the Fermi and Gamow-Teller distributions with the appropriate weighting factors. The program takes care of this automatically when providing a mixing ratio as user input. The effect of the recoil after $\beta$ decay of a nucleus of finite mass $M$ is to multiply the phase space by a factor $R_N(W, W_0, M)$. As the recoiling nucleus carries away energy, this decreases the available phase space for the outgoing leptons.

\item \textit{$Q(Z, W)$: }
A final consequence of the finite nuclear mass and consequent recoil is a change in the Coulomb field in which the departing electron or positron moves. It is not fixed in space but is itself recoiling against the combined lepton momenta so that the field experienced by the $\beta$ particle differs with time from what it would have been if the nucleus would not be recoiling, as is assumed for the Fermi function.

\item \textit{$R(Z, W, W_0)$: }
The interaction of the ejected $\beta$ particle with the charged nucleus has to be calculated to several orders of perturbation theory. This is already partly taken care of through the inclusion of the Fermi function. Higher-order loop corrections are not included, however. The additional radiative corrections which we deal with here are energy- and nucleus dependent, and are typically referred to as the ``outer'' radiative corrections. They correspond to the exchange of virtual photons or $Z^0$ bosons between the charged particles involved in $\beta$ decay, and are calculated up to order $Z^2\alpha^3$. An distinction is made between $\beta$ particle and (anti)neutrino radiative corrections, where the latter are calculated according to Sirlin \cite{Sirlin2011}.

\item \textit{$S(Z, W)$: }
The nucleus cannot be completely separated from its orbiting electrons as the decay is governed by the total Hamiltonian. Even though the interaction point lies within the nucleus, the emitted $\beta$ particle undergoes continuous interaction with the atomic electrons that surround it. The electric potential at the $\beta$ creation site is more shallow because of a non-zero probability of finding atomic electrons inside the nucleus. This introduces a screening correction, as the effective nuclear charge that the $\beta$ particle sees is lowered, which in turn lowers (increases) the decay rate for $\beta^-$ ($\beta^+$) decay. When considering the total molecular system, the situation becomes increasingly complex as we now have to take into account nearby nuclei and the total, shared electron distribution. Analytic estimates have been given in Ref. \cite{Hayen2018} that are, however, not implemented in this code due to its case-by-case nature. 

\item \textit{$X(Z, W)$: }
The non-orthogonality of initial and final state atomic wave functions in $\beta$ decay allows for additional indirect processes through which electrons can be emitted into a continuum state. In case of the exchange effect, this non-orthogonality leaves a possibility for a $\beta$ particle to be emitted directly into a bound state of the daughter atom, thereby expelling an initially bound electron into the continuum. As both electrons are indistinguishable, this gives a positive contribution to the decay rate. The effect is significant for low energies, but becomes negligible typically on the 30 to 200 keV range for low and high $Z$ nuclei, respectively.

\item \textit{$r(Z, W)$: }
The $\beta$ decay of a nucleus results in a sudden change of the nuclear potential, both due to a charge difference as well as a recoil effect. As the eigenstates for initial and final states belong to slightly different Hamiltonians, the initial and final atomic orbital wave functions only partially overlap. This allows for discrete processes such as shake-off and shake-up, which decrease the decay rate as the phase space volume is reduced. The so-called atomic mismatch correction takes this into account in an average way. The code allows the user to instead specify an exact energy deficit due to atomic excitations, thereby replacing this contribution.

\item \textit{$C(Z, W)$: }
All nuclear structure-sensitive information and radial wave function behaviour is combined into the so-called shape factor, $C$. The former is concerned with four nuclear form factors relevant to allowed $\beta$ decay, the precise evaluation of which will be discussed in greater detail in Sec. \ref{sec:nuclear_structure}. The latter deals with both the decrease of the radial leptonic wave functions inside the nuclear volume and the shape of the \textit{weak} charge density, denoting the volume obtained from initial and final nuclear wave functions participating in the decay\footnote{A thorough discussion of the use of `finite size' corrections in the literature can be found in the original work \cite{Hayen2018}.}. The required nuclear form factors are free parameters, to be obtained either through some type of many-body calculation or through the Conserved Vector Current hypothesis when appropriate. Here, the weak charge density is approximated as being the charge density. For high precision results this requires an additional correction.


\item \textit{$C_I(Z, W)$: }
As mentioned in the previous correction, the weak charge density is initially approximated to be the same as the nuclear charge density. In reality, however, the former is built from the initial and final participating nuclear states, which will differ (significantly) from the pure charge distribution due to angular momentum conservation and phase space volume. A correction in an extreme single-particle fashion, where shells are filled in the regular $jj$-coupling fashion, is then applied. In this code it is possible to use the numerical results of single-particle wave functions in the calculation of this correction factor.

\item \textit{$D_{\text{FS}/C}(Z, W, \beta_2)$: }
Several regions of the nuclear chart contain (heavily) deformed nuclei, often with shape coexistence nearby and a rapid change between prolate and oblate distributions. Changing a spherical uniformly charged density to an ellipsoidal form changes the $L_0$ correction described above by $D_{\text{FS}}$, as well as the convolution of the lepton and nucleon wave functions described in the $C_I$ correction written as $D_C$. This is generally a small effect, but increases linearly with $Z$ and so can become significant on the few $10^{-4}$ level in the fission fragment and lead region.

\end{itemize}
All of these corrections can be turned off individually in the code in order to investigate the influence and associated uncertainties of each of them.

\section{Nuclear matrix element evaluation}
\label{sec:nuclear_structure}
As nuclear $\beta$ is very much a \textit{nuclear} process, some part of the transition amplitude will depend on nuclear structure information. In a popular approach both initial and final nuclei are considered elementary particles, and this nuclear structure information is encoded in form factors categorised through conservation of angular momentum. These are dynamical quantities depending on the four-momentum transfer in the reaction, $q^\mu$. Given the small amplitude of $q$ in regular $\beta$ decay, a useful consequence of this approach allows for a perturbation calculation by expanding the form factors in terms of the available Lorentz scalar, $q^2$. Except for the main Fermi or Gamow-Teller form factors, it is typically sufficient to retain only the constant term for $q^2=0$. Following the results in standard works by Behrens and collaborators \cite{Behrens1982} and Holstein \cite{Holstein1974}, the mathematical rigour of the former was combined with the clear notation and symmetry properties of the latter when constructing the dominant $C(Z, W)$ correction \cite{Hayen2018}.

In theory one could stop here and determine the relevant form factors experimentally. One can, however, attempt to calculate these form factors using some type of nuclear many-body calculation. Following some approximations, form factors are then transformed into nuclear matrix elements. Here, we briefly summarise the structure and evaluation of these matrix elements.

\subsection{General formulation}
A common approach to reduce nuclear form factors to matrix elements is performed by introducing the impulse approximation. The latter states that the nuclear current can be considered as the sum of individual nucleon currents which behave as if they were free particles. In doing this, one neglects meson-exchange and off-mass shell corrections. As a concession for this simple approximation, a quenching of the coupling constants that dominate the decay is introduced. In the case of nuclear $\beta$ decay, the axial vector coupling constant, $g_A$, is typically reduced to values below unity compared to usual many-body calculations \cite{Suhonen2017a}. The approximation allows, however, a transformation of the form factors into nuclear one-body \textit{matrix elements}. In order to do this, we write the operators in second quantisation. For a general operator of rank $L$ in spherical tensor form, $O_{L}^M$, we write
\begin{equation}
\langle f | O_{L}^M \tau^{\pm} | i \rangle = \sum_{\alpha, \beta} \langle \alpha | O_{L}^M | \beta \rangle \langle f | a^\dagger_\alpha a_\beta | i \rangle
\label{eq:operator_decomp}
\end{equation}
where $\alpha$ and $\beta$ are single-particle proton (neutron) and neutron (proton) states for $\beta^-$ ($\beta^+$) decay. As one is usually interested only in the reduced matrix elements, one writes
\begin{equation}
    \langle f || \mathbf{O}_L || i \rangle = \hat{\lambda}^{-1} \sum_{\alpha \beta} \langle \alpha || \mathbf{O}_L || \beta \rangle \langle f || [a^\dagger_\alpha \tilde{a}_\beta]_\lambda || i \rangle,
    \label{eq:reduced_matrix_elements_decomposition}
\end{equation}
where $\hat{\lambda} = \sqrt{2\lambda + 1}$ and $[a^\dagger_\alpha \tilde{a}_\beta]_\lambda$ represents the coupling of the creation and annihilation spherical tensors to a total angular momentum $\lambda$. This quantity is typically referred to as the reduced one body transition density (ROBTD), and is readily calculated by nuclear shell model codes such as NuShellX@MSU \cite{Brown2014}. Given these inputs, it remains to calculate the single particle reduced matrix elements $\langle \alpha || \mathbf{O}_L || \beta \rangle$. This is the topic of the next section. 


\subsection{Single-particle matrix elements}
Due to their supreme importance, we first discuss the evaluation of the single-particle reduced matrix elements, $\langle \alpha || \mathbf{O}_{L} || \beta \rangle$. The general solution to the Dirac equation in a spherical potential can be written as \cite{Behrens1982}
\begin{equation}
\Psi_{\kappa} = \left(
\begin{array}{c}
\text{sign}(\kappa)f_{\kappa}(r) \sum_{\mu}\chi_{-\kappa}^{\mu} \\
g_{\kappa}(r) \sum_{\mu}\chi_{\kappa}^{\mu}
\end{array}
\right),
\label{eq:sol_dirac}
\end{equation}
where $\kappa$ is the eigenvalue of the usual $\hat{K}=\beta\{(\bm{\sigma}\cdot \bm{L})+1 \}$ operator, $\mu$ is the projection of the total angular momentum on the $z$-axis, and $f(r)$ and $g(r)$ are the small and large radial solutions, respectively. Through coupling of initial and final states, matrix elements of varying magnitude are constructed. The smallest of these, so-called relativistic, matrix elements couple the small radial functions of both initial and final states and can typically safely be neglected. In this work, the remaining terms are treated non-relativistically where for cross terms the following reduction is used
\begin{equation}
f_\kappa(r) = \frac{1}{2M_N}\left(\frac{d}{dr}+\frac{\kappa+1}{r} \right)g_\kappa(r),
\end{equation}
with $M_N$ the nucleon mass. Using this one obtains a radial Schr\"odinger equation for $g_\kappa(r)$.

In order to properly evaluate the single-particle matrix elements, one has to then accurately know the radial wave function of initial and final states. Due to their simplicity we write these in a spherical harmonic oscillator basis, with the additional benefit of analytical matrix elements. The independent particle solution in a spherical potential can then be written as a linear combination of spherical harmonic oscillator wave functions
\begin{equation}
|\nu j \rangle = \sum_N C_{Nj}^\nu |N j \rangle
\label{eq:wf_spherical_exp}
\end{equation}
with $N$ the oscillator quantum number and $j$ the total angular momentum. As discussed by Refs. \cite{Behrens1970, Hayen2018}, all operator evaluations contain terms $\int g_i(r)\, (r/R)^M g_f(r) r^2 dr$, with integer $M$. After resolving angular momentum coupling, one has to evaluate matrix elements of the type $\langle N_f l_f | r^M | N_i l_i \rangle$, for which we use the closed formula by Nilsson \cite{Nilsson1955}. Single particle matrix elements are then obtained by simply summing over all harmonic oscillator contributions.

While previous results are useful in spherical nuclei, often one is confronted with strong nuclear deformation. In this case, the orbital quantum number $j$ is no longer conserved, and one considers instead the projection of $\bm{j}$ along the symmetry axis, denoted by $\Omega$. Writing the angular momentum of a rotating deformed core as $\bm{R}$, we denote by $K$ the projection along the symmetry axis of the sum $\bm{j}+\bm{R}$. The new single-particle wave function will then be a combination of wave functions of the type of Eq. (\ref{eq:wf_spherical_exp}) with the appropriate spin projections such that
\begin{equation}
| \mu \Omega \rangle = \sum_{Nj} \sum_\nu C^\mu_{\nu \Omega} C^\nu_{Nj} | N j \rangle \equiv \sum_{j} C_{j\Omega} | N l j \Omega \rangle
\label{eq:wf_deformed_exp}
\end{equation}
where in the last term we have written our solution using the notation by Davidson\footnote{The difference with the treatment of Davidson lies in the spherical potential used. Unless this is exactly a modified spherical oscillator wave function as per Nilsson \cite{Nilsson1955}, more than one $C^\nu_{Nj}$ will be non-zero. In the limit of zero deformation, Eq. (\ref{eq:wf_spherical_exp}) will generally contain more than one term.} \cite{Davidson1968}. In the calculation $\Omega$ is defined as a positive number. A negative projection is obtained through the relation
\begin{equation}
C_{j\Omega} = (-1)^{\frac{1}{2}-j}\pi_\Omega C_{j-\Omega}
\end{equation}
with $\pi_\Omega = (-1)^l$ the parity of the state. The coefficients $C_{j\Omega}$ are calculated numerically and are normalised according to $\sum_j |C_{j\Omega}|^2 = 1$. The evaluation of reduced single particle matrix elements in the deformed case is summarized for clarity in the appendix.

\subsection{Form factors in allowed decay}
The preceding sections discussed the evaluation of general form factors using the impulse approximation in a general sense. In allowed $\beta$ decay, however, only a limited number of form factors are to be calculated. The calculations implemented in this work are based on the formalism by Behrens and B\"uhring \cite{Behrens1982}, where form factors are denoted as $F_{KLs}(q^2)$. Different orders of their expansion in $q^2$ are written as $F^{(n)}_{KLs}$, and can be evaluated using the impulse approximation. Table \ref{table:matrix_elements_BB} shows the form factor evaluation in the impulse approximation and its relation to the well-known Holstein form factors for allowed Gamow-Teller decay.

\begin{table}[ht]
\centering
\caption{Overview of the relevant form factors for Gamow-Teller decay which need to be evaluated in a nuclear structure sensitive way. The first column shows the form factor notation by Behrens and B\"uhring (BB), while the second shows the relation of the former to the notation of Holstein (HS). Here, $M$, $M_N$ and $R$ are the nuclear mass, nucleon mass and nuclear radius, respectively. Finally, the last column shows the evaluation of the form factor in the impulse approximation in cartesian coordinates. Here the explicit integration over nucleon degrees of freedom together with the isospin ladder operators are left out for notational convenience. All form factors are identically equal to zero when considering $0 \to 0$ transitions since all operators are rank 1 tensors in spin space. The coupling constants for the axial, weak magnetism, and induced pseudoscalar currents are written as $g_A$, $g_M$ and $g_P$, respectively. Finally, $\bm{\sigma}$ is the four-dimensional equivalent of the Pauli matrices, $\beta= -\gamma_0$ and $\bm{\alpha} = \gamma_5 \bm{\sigma}$ using the standard Dirac $\gamma$-matrices.}
{\renewcommand{\arraystretch}{1.9}
\begin{tabular}{cllcc}
\hline \hline
Form Factor (BB) & Form Factor (HS) & Matrix element (Impulse Approximation) \\
\hline
$^AF_{101}^{(0)}$ & $- c_1$ & $\mp g_A\int \bm{\sigma}$ \\
$^AF_{121}^{(0)}$ & $-\frac{5\sqrt{2}}{4}\frac{1}{(MR)^2}h$ & $\mp g_A \frac{3}{\sqrt{2}}\int \frac{(\bm{\sigma}\cdot \bm{r})\bm{r}-\frac{1}{3}\bm{\sigma}\cdot r^2}{R^2} \mp g_P\frac{2M_N}{R}5\sqrt{2}\int \beta \gamma_5 \frac{i \bm{r}}{R}$\\
$^VF_{111}^{(0)}$ & $-\sqrt{\frac{3}{2}}\frac{1}{MR}b$ & $-g_V\sqrt{\frac{3}{2}}\int \frac{\bm{\alpha} \times \bm{r}}{R} - (g_M-g_V)\frac{2M_N}{R}\sqrt{6}\int \bm{\sigma}$ \\
$^AF_{110}^{(0)}$ & $-\frac{\sqrt{3}}{2MR}d$ & $\mp g_A \sqrt{3}\int \gamma_5 \frac{i\bm{r}}{R}$\\
\hline \hline
\end{tabular}
}
\label{table:matrix_elements_BB}
\end{table}
As described in the preceding sections, the matrix elements are evaluated in the non-relativistic approximation. In a spherical harmonic oscillator basis, all single-particle matrix elements are analytical, such that only the coefficients, $C_{j\Omega}$ need to be determined. Combined with ROBTD from a nuclear many-body calculation, the full reduced matrix elements of Eq. (\ref{eq:reduced_matrix_elements_decomposition}) can be calculated to high precision. 

\section{Beta Spectrum Generator: Options}
\label{sec:BSG_options}
\begin{figure}[!ht]
\centering
\includegraphics[width=0.5\textwidth]{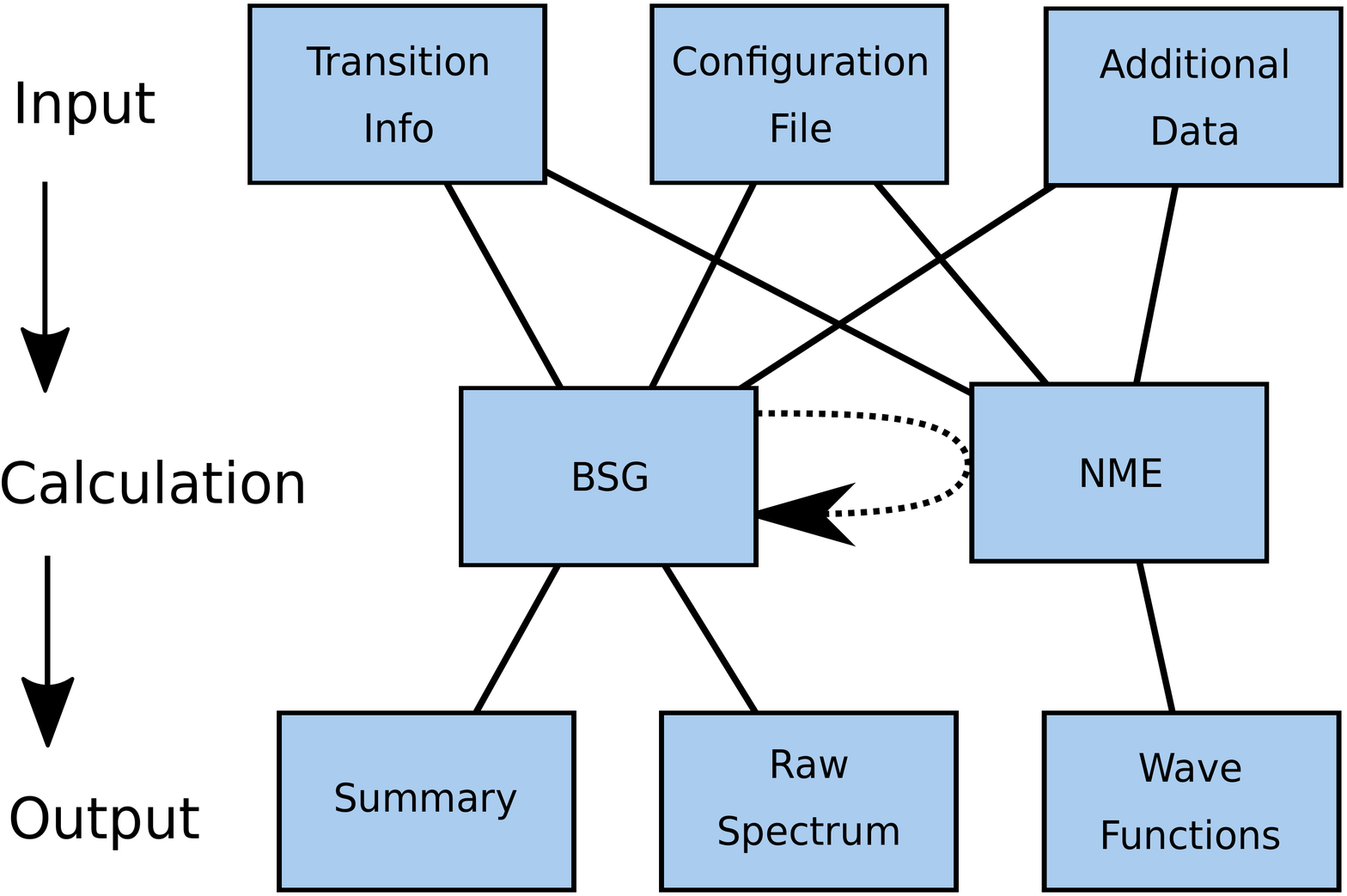}
\caption{Flowchart describing the general structure of the libraries, their required inputs and outputs. The dotted arrow from \texttt{BSG} to \texttt{NME} signifies the possible connection between the two libraries when matrix elements are not provided by the user or proper single-particle wave functions are used in the corrections sensitive to the weak charge distribution.}
\label{fig:flowchart}
\end{figure}

For a more detailed explanation the reader is referred to the online documentation and the in-line documentation. Likewise, for an user guide to the graphical user interface, the reader is referred to the online documentation. We present here a brief overview of the typical usage and additional possibilities of the package. A flowchart of the structure is shown in Fig. \ref{fig:flowchart}.

\subsection{General usage}
In its simplest mode of operation, the \texttt{BSG} code requires a single configuration file for general properties and one transition-specific file. Both are written in typical INI-style, examples of which are provided with the software distribution. In the absence of nuclear structure calculations, the configuration file simply contains the following options:
\begin{itemize}
\item \texttt{Constants.gA}: The value for the axial vector coupling constant.
\item \texttt{Constants.gP}: The value for the induced pseudoscalar coupling constant, relative to the nucleon rather than muon mass. See Ref. \cite{Hayen2018} for details.
\item \texttt{Constants.gM}: The value for the weak magnetism coupling constant.
\end{itemize} 
Here the dot separtes the header from the variable name in usual INI fashion.
The transition-specific file contains the following options:
\begin{itemize}
\item \texttt{Transition.Process}: The process of the $\beta$ transition. Possibilities are \texttt{B-} ($\beta^-$), \texttt{B+} ($\beta^+$) and \texttt{EC} (electron-capture).
\item \texttt{Transition.Type}: The type of the decay. Possibilities are \texttt{Fermi}, \texttt{Gamow-Teller} and \texttt{Mixed}.
\item \texttt{Transition.MixingRatio}: The mixing ratio between Fermi and Gamow-Teller components, defined as $\rho = g_AM_{GT}/g_VM_F$. Only relevant when the type is set to \texttt{Mixed}.
\item \texttt{Transition.QValue}: The $Q$-value of the decay as the difference in atomic masses between initial and final \textit{ground} states. Units are in keV.
\item \texttt{Transition.PartialHalflife}: The partial half-life of the transition in seconds, defined as $t_{1/2}^p = (t_{1/2}/\text{BR})[1+\epsilon/\beta^+]$, where BR is the branching ratio and $\epsilon$ the electron-capture probability relative to that of $\beta^+$ decay. This is used to calculate the theoretical $ft$ value.
\item \texttt{Transition.AtomicEnergyDeficit}: In case the energy deficit to the endpoint energy, $\Delta W_0$, due to atomic final state excitations is known explicitly the user can override the $r(Z,W)$ correction and use this value instead.
\item \texttt{Transition.ROBTDFile}: Externally calculated reduced one body transition densitites, organised according to the angular momentum coupling as in Eq. (\ref{eq:reduced_matrix_elements_decomposition}). Explicit support is provided for \texttt{.obd} files resulting from NuShellX@MSU calculations \cite{Brown2014}. See Sec. \ref{sec:NME_options} for a more elaborate discussion.
\end{itemize}
Finally, information must be provided about the initial and final states of the $\beta$ decay of interest. This is performed according to the following options, with headers \texttt{Mother} and \texttt{Daughter}\footnote{Deformation parameters can be deduced from standard methods, and are calculated theoretically by M\"oller \emph{et al.} \cite{Moller2016} for a large series of isotopes. Electric quadrupole moments are collected e.g. by Stone \cite{Stone2005}. In case the nuclear spin is 0 or $1/2$, the spectroscopic quadrupole moment is zero, and possible deformation can be experimentally deduced from rotational bands or charge radii.  Nuclear charge radii have been collected by Angeli and Marinova \cite{Angeli2013} and extended by Bao \emph{et al.} \cite{Bao2016}.}:
\begin{itemize}
\item \texttt{Z}: The proton number of the state.
\item \texttt{A}: The mass number of the state.
\item \texttt{Radius}: The nuclear charge radius in fm. When absent, the Elton formula \cite{Elton1958} is used.
\item \texttt{SpinParity}: A string denoting the double of the angular momentum multiplied by the parity of the level. As an example, \texttt{-7} represents a $7/2^-$ state.
\item \texttt{Beta2-6}: Different nuclear deformation parameters as defined in Eq. (\ref{eq:nuclear_surface_deformation}).
\item \texttt{ExcitationEnergy}: Excitation energy of the state in keV. This value is combined with the $Q$-value of the decay to obtain the correct endpoint energy.
\end{itemize}
The configuration and transition file names can be fed into the program using the \texttt{-c} and \texttt{-i} flags, respectively. 

\subsection{Extended options}
By default, all corrections described in Sec. \ref{beta-full} are turned on. If so required, the user can turn each of these off individually through flags on the command line or additional options in the configuration file under the \texttt{Spectrum} header. More information can be obtained by running the program with the \texttt{--help} flag.

While all corrections described in the previous section have been derived analytically, there is some remaining model dependency which can be selected by the user. This deals primarily with the choice of the nuclear charge distribution. Possibilities here include
\begin{itemize}
\item \texttt{Spectrum.ESShape}: Denotes the more precise nuclear charge distribution to be used in the electrostatic shape correction factor, $U$. The implemented options are \texttt{Fermi}, \texttt{Modified Gaussian} and \texttt{Custom}. In the case of the Fermi distribution, a parametrization as a function of the proton number was given by Wilkinson \cite{Wilkinson1993b}, while the Modified Gaussian result uses the results by obtained Hayen \textit{et al.} \cite{Hayen2018} after fitting the composite charge distribution according to Ref. \cite{Wilkinson1993b}. The \texttt{Custom} possibility allows the user to specify the $v_{0,2,4}^{(')}$ potential expansion coefficients of both reference and new electrostatic potentials according to Ref. \cite{Hayen2018}.
\item \texttt{Spectrum.NSShape}: Allows the user to choose which nuclear charge distribution is to be used in the $C$ correction. By default, a modified Gaussian distribution is employed, obtained using a fit through a composite charge distribution built up out of harmonic oscillator wave functions \cite{Wilkinson1993b}.
\end{itemize}
Finally, the code allows the user to change the $C_I$ correction through the option
\begin{itemize}
\item \texttt{Spectrum.Coupled}: Couples the \texttt{BSG} and \texttt{NME} libraries to use numerically calculated single-particle wave functions of initial and final states as a linear combination of spherical harmonic oscillator wave functions instead of the simplest $jj$-coupling approach.
\end{itemize}

\subsection{Nuclear matrix elements}
The required nuclear matrix elements are free parameters in the analytical formulation of the shape factor, $C$. As such they have to be provided for the calculation to be able to run. The easiest way of doing this is through manual specification on the command line. This is performed using the \texttt{-b}, \texttt{-d} and \texttt{-L} flags, representing the weak magnetism ($b/Ac$), induced tensor ($d/Ac$) and $\Lambda$ corrections, respectively. These quantities are related to the form factor notation by Behrens and B\"uhring through
\begin{align}
    \frac{b}{Ac} &= \frac{1}{g_A}\left((g_M - g_V) -M_NR\sqrt{\frac{2}{3}}\frac{^VF_{111}^{(0)}}{^AF_{101}^{(0)}}\right) \label{eq:bAc_BB}\\
    \frac{d}{Ac} &= M_NR\frac{2}{\sqrt{3}}\frac{^AF_{110}^{(0)}}{^AF_{101}^{(0)}} \label{eq:dAc_BB}\\
    \Lambda &= \frac{^AF_{121}^{(0)}}{^AF_{101}^{(0)}} \label{eq:Lambda_BB}
\end{align}
where $M_N$ is the nucleon mass. A more extended explanation concerning the origins of these terms can be found in Ref. \cite{Hayen2018}.

\section{Nuclear matrix elements: Options}
As in the previous section, the reader is referred to the online documentation and the in-line documentation collected by Doxygen for a more in-depth explanation. We present here a brief overview of the typical usage and additional possibilities of the package.
\label{sec:NME_options}
\subsection{General options}
The operation of the \texttt{NME} library is simple in its use. The input file for a specific transition requires the same type of information as the \texttt{BSG} library concerning the physical properties of initial and final nuclear states. The library enables automatic calculation of the weak magnetism and induced tensor corrections, normalised by $Ac_1$, using the following flags
\begin{itemize}
\item \texttt{-b [--weakmagnetism]}: Calculates the weak magnetism contribution as $b/Ac_1$, i.e. normalized with the mass number and Gamow-Teller matrix element.
\item \texttt{-d [--inducedtensor]}: Calculates the induced tensor contribution as $d/Ac_1$, normalized just like the weak magnetism contribution.
\item \texttt{-M Wxyz}: Calculates a specific matrix element $^W\mathcal{M}_{xyz}$, where $W$ can be $V$ or $A$ as in the Behrens-B\"uhring description.
\end{itemize}
These options are called automatically in the \texttt{BSG} library when specific values are not provided. Except for the coupling constants defined in the previous section, the \texttt{NME} allows for a quenching of $g_A$, by defining the \texttt{Computational.gAeff} constant in the configuration file.

For the nuclear structure component two options are currently provided. The first is the inclusion of externally calculated reduced one body transition densities, originating for example from the nuclear shell model. The second is the built-in and default option of the extreme single-particle evaluation. The choice is determined by the following parameter
\begin{itemize}
    \item \texttt{Computational.Method}: Determines the method of evaluating the matrix elements. Setting it to \texttt{ROBTD} requires an external file specified in the transition info, while \texttt{ESP} attempts to calculate the matrix elements in an extreme single-particle fashion.
\end{itemize}

\subsection{External ROBTD calculations}
\label{sec:advanced_methods}
Section \ref{sec:nuclear_structure} discussed the nature of the nuclear form factors required for the allowed $\beta$ spectrum shape. Its evaluation is based on the impulse approximation and assuming the interaction occurs between the particles closest to the Fermi energy. While this is sometimes a good approximation for low to medium $Z$, odd-$A$ nuclei, the validity for higher masses and even-$A$ $\beta$ transitions is expected to break down. Transitions to or from excited states typically cannot be understood in a purely single-particle manner. In mid-shell regions, even the ground state is often a complex mixture of valence particle interaction and collectivity features. One requires then a more elaborate method of calculating these nuclear matrix elements. Typically, this is done using either the nuclear shell model or mean field theories. As mentioned at the beginning of this section, this can be connected to the single-particle solutions by writing the isospin ladder operator in second quantisation as in Eq. (\ref{eq:operator_decomp}). A sum is made over all single-particle states $\alpha$, $\beta$ such that their coupled angular momenta satisfy the triangle relation with initial and final nuclear spins. The code presented here allows for input of the OBDME without prior requirements on its origin. This is done by adding an additional option to the transition-specific file under the \texttt{Transition} header
\begin{itemize}
\item \texttt{ROBTDFile}: Sets the location of the file containing the one-body density matrix elements in a spherical harmonic oscillator basis. This is a CSV-type file where each line is written as `$\lambda, 2\cdot j_\beta, n_\beta, l_\beta, 2\cdot j_\alpha, n_\alpha, l_\alpha,\langle f || [a^\dagger_\alpha \tilde{a}_\beta]_\lambda || i \rangle$', where $\lambda$ is the angular momentum coupling of creation and annihilation operators as before. Note that explicit support is provided for \texttt{.obd} files coming from a NuShellX@MSU calculation\footnote{In NuShellX@MSU, the reduced one body transition densities are reported as $\hat{\lambda}^{-1}\langle f || [a^\dagger_\alpha \tilde{a}_\beta]_\lambda || i \rangle$, i.e. the factor $\hat{\lambda}^{-1}$ is already accounted for. This is taken into account when loading the \texttt{.obd} results.}.
\end{itemize}
The full reduced matrix elements are then calculated using the analytical results for harmonic oscillator single particle reduced matrix elements.


\subsection{Extreme Simple Particle Evaluation}
As mentioned in preceding sections, the \texttt{NME} library will attempt to evaluate nuclear matrix elements in an extreme single-particle manner when no external input is given. The essential ingredient is then the description of initial and final single-particle states. The code described here implements different ways of evaluating the matrix elements of Table \ref{table:matrix_elements_BB} in a single-particle manner. In all approaches, nuclear wave functions are expressed as a linear combination of spherical harmonic oscillator wave functions for which the single-particle matrix elements are analytically calculated. It remains then to calculate the $C_{Nj}^\nu$ (Eq. (\ref{eq:wf_spherical_exp})) or $C_{j\Omega}$ (Eq. (\ref{eq:wf_deformed_exp})) in spherical and deformed cases, respectively. 
Currently three options for the potential are available, specified through the \texttt{Computational.Potential} option in the configuration file. For all but the first method, the code utilises a custom C++ port of the work by Hird \cite{Hird1973}.

\begin{itemize}
\item \texttt{SHO}: The simplest approach of all fills the nuclear potential in a typical $jj$-coupling manner. Protons and neutrons are thus filled according to the $1s, 2s, 1p_{3/2}, 1p_{1/2}, \ldots$ filling scheme, and the resultant final particle is approximated by a harmonic oscillator wave function with primary and orbital quantum numbers $n$ and $l$ of the final orbital. This includes no deformation, and is charge-insensitive.
\item \texttt{WS}: Wave functions are solved in a spherical Woods-Saxon potential. The Hamiltonian is constructed in a spherical harmonic oscillator basis with variable basis size up to primary quantum number $n=12$ or $n=13$ for parity-even and parity-odd states, respectively. The resultant Hamiltonian is diagonalised and yields eigenvalues and eigenvectors for all bound states. The eigenstates are filled and a set of $C_{Nj}^\nu$ are returned for the final nucleon.
\item \texttt{DWS}: Wave functions are solved in an axially deformed Woods-Saxon potential, which includes quadrupole, $\beta_2$, hexadecupole, $\beta_4$, and dotriacontupole, $\beta_6$, deformations. The approach is done analogously to that of the former and one obtains a set of eigenvalues and eigenvectors in a spherical harmonic oscillator basis. 
\end{itemize}
The Hamiltonian used for this endeavour is based on an extended Woods-Saxon potential \cite{Hird1973}
\begin{equation}
\mathcal{H} = -\frac{\hbar^2}{2m}\nabla^2 - V_0f(r) - V_s\left(\frac{\hbar}{m_\pi c}\right)^2\frac{1}{r}\frac{df}{dr}\bm{l}\cdot\bm{s} + V_0R\frac{df}{dr}\sum^6_{n\text{ even}}(\beta_{n}Y^0_n)
\end{equation}
where $V_0$ is the depth of the Woods-Saxon potential, $f(r) = (1+\exp[(r-R)/a_0])^{-1}$ is the Woods-Saxon function with surface thickness $a_0$ and nuclear radius $R$, and spherical harmonics, $Y_{L}^M$. The spin-orbit term contains the coupling constant, $V_s$, and the Compton wavelength of the pion, $\hbar/m_\pi c$. In the case of protons, an additional Coulomb potential is added. The depth of the Woods-Saxon potential is given in its `optimized' form as \cite{Dudek1982, Dudek1980}
\begin{equation}
V_0 = V\left(1\pm \chi \frac{N-Z}{N+Z}  \right)
\end{equation}
with the upper (lower) sign for protons (neutrons), and $V=-49.6\,$MeV and $\chi=0.86$ by default. All parameters can be specified in the configuration file through the following options under the \texttt{Computational} header in case the default values do not suffice
\begin{itemize}
\item \texttt{SurfaceThickness}: Sets the surface thickness, $a_0$, of the Woods-Saxon potential in femtometer.
\item \texttt{Vneutron/Vproton}: Sets the base depth of the Woods-Saxon potential in MeV for neutron and proton states.
\item \texttt{Xneutron/Xproton}: Sets the potential asymmetry for neutron and proton states.
\item \texttt{V0Sneutron/V0Sproton}: Sets the spin-orbit coupling constant for neutron and proton states.
\end{itemize}

As the simple filling schemes based on these potentials do not always accurately predict the correct valence spin state, the code allows for an enforcement of the correct single-particle states within a user-specified energy margin. It contains several options to enforce spin selection and coupling
\begin{itemize}
\item \texttt{ForceSpin}: Instead of choosing the single-particle state that is obtained through simple filling, pick the closest one corresponding to the spin state defined in the \texttt{Mother/Daughter.ForcedSPSpin} in even-$A$ nuclei. In odd-$A$ nuclei, the correct spin state is automatically chosen if one can be found within the user-defined energy window.
\item \texttt{ReversedGallagher}: Most examples of deformed, even-$A$ nuclei follow the simple spin selection rules by Gallagher \cite{Gallagher1958}. In the original work, a `reversed' selection rule was defined, which can be turned on in the code.
\item \texttt{OverrideSPCoupling}: In case a properly coupled state cannot be obtained using conventional coupling rules, override the coupling completely and set the coupled spins to the corresponding nuclear state.
\item \texttt{EnergyMargin}: Set the margin in MeV to select a different state corresponding to the proper initial or final state when it is not obtained through regular methods.
\end{itemize} 
When activated, the state closest in energy to the originally proposed level with the correct spin state is selected as the one that participates in the interaction. 

\section{Example test cases}
\label{sec:usage_examples}
\subsection{\texttt{NME}}
An ideal candidate for testing the performance of the calculations performed by the \texttt{NME} part of the code is the weak magnetism form factor of mirror $\beta$ transitions, as its exact value can be calculated from experimental magnetic dipole moments as follows \cite{Holstein1974}
\begin{equation}
b^\mp = \mp A \sqrt{\frac{J+1}{J}} (\mu_M-\mu_D)
\label{eq:b_CVC}
\end{equation}
with $J$ the total angular momentum of initial and final mirror states, $A$ the mass number and $\mu_M$ ($\mu_D$) the magnetic dipole moment of the mother (daughter). Due to its presence in the spectrum shape, Eq. (\ref{eq:b_CVC}) is often divided by $Ac_1$ to obtain an experimental ratio. The leading order Gamow-Teller form factor, $c_1$, is obtained from the $ft$ value. In terms of nuclear matrix elements this ratio becomes
\begin{equation}
\frac{b}{Ac_1} = \frac{1}{g_A}\left(g_M + g_V \frac{M_L}{M_{GT}} \right)
\end{equation}
where the matrix elements definitions follow those by Holstein \cite{Holstein1974}. 

\subsubsection{Weak magnetism in the extreme single-particle evaluation}
As was mentioned in Sec. \ref{sec:nuclear_structure}, we have to deal here with a \textit{ratio} of matrix elements of the same order in spherical tensor formulation, meaning complex many-body couplings drop out in the absence of core polarisation and meson exchange. We are left, then, with a ratio of single-particle matrix elements such that we expect the extreme single-particle to capture most of the required dynamics. The challenge then remains to pick a suitable single-particle state based on an underlying potential. 

\begin{figure}[!ht]
\centering
\includegraphics[width=0.48\textwidth]{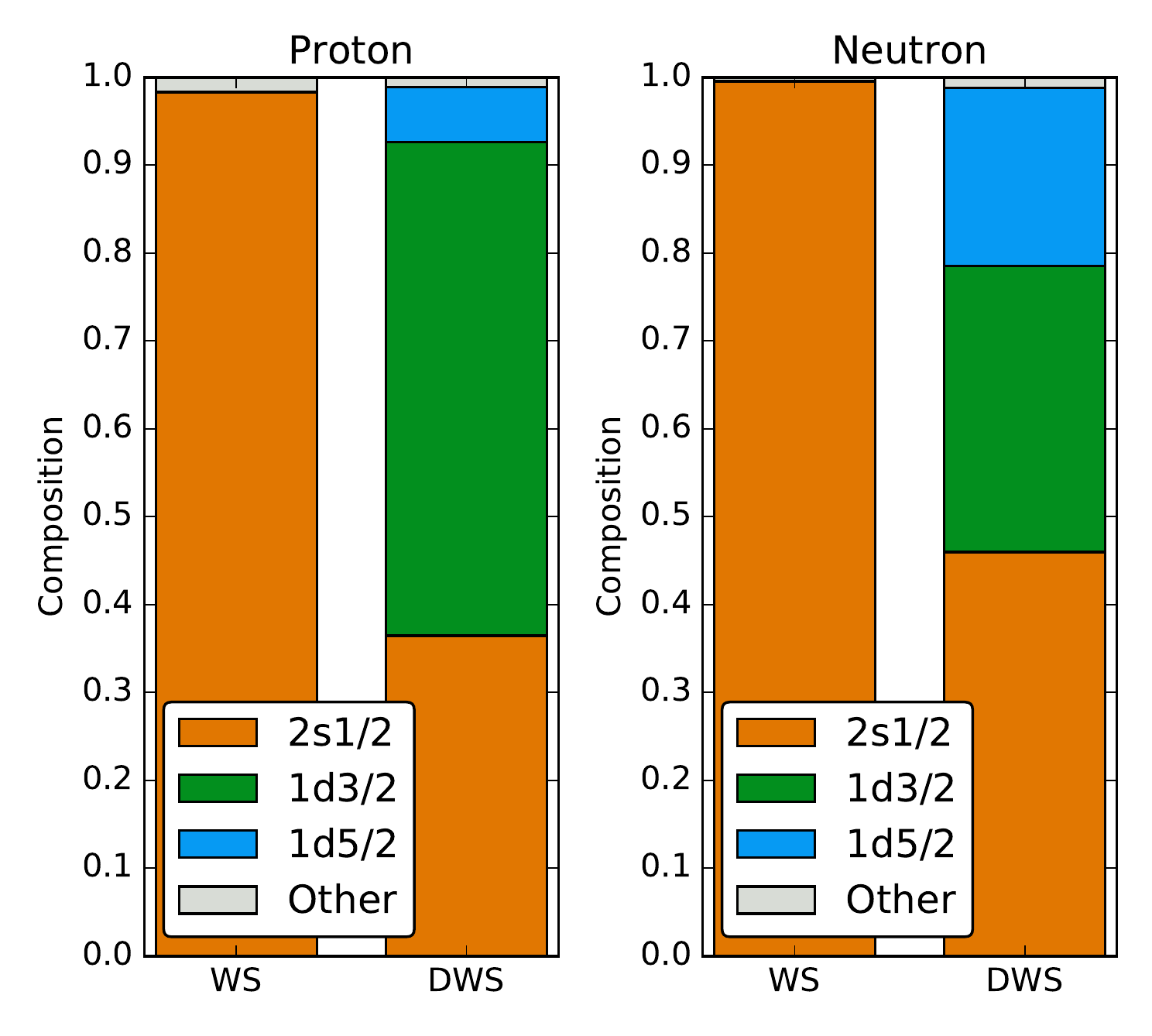}
\caption{Single particle wave function composition for the spherical Woods-Saxon (WS) and deformed Woods-Saxon (DWS) potentials for both proton (initial) and neutron (final) states of the $^{31}$S $\beta^+$ mirror decay. Large admixtures of higher-spin orbitals are expected in the deformed potential due to the 1/2$^+$ initial and final spin, as all orbitals contain $\Omega = 1/2$ projections. Results for weak magnetism remain unchanged, however, with values equal to 5.19 and 5.18 for spherical and deformed potentials, respectively.}
\label{fig:wf_comp_31S}
\end{figure}

\paragraph{$^{31}$S}
As a first example, we consider the $\beta^+$ mirror decay of $^{31}$S for which the experimental $b/Ac_1$ value can be calculated to give $5.351(14)$. The spin of initial and final state can be understood from a simple $jj$-filling scheme based on the standard shell model orbitals, and indeed choosing the simple spherical harmonic oscillator potential results in a calculated $b/Ac_1$ value of 5.19 in agreement with the experimental value (obtained from the mother and daughter magnetic moments and the $^{31}$S mirror $\beta$ transition $ft$ value) within 3\%. Moving to the spherical Woods-Saxon potential leaves this value unchanged, as the wave function is completely dominated by the $2s_{1/2}$ orbital. According to the mean field results by M\"oller \emph{et al.} \cite{Moller2016}, however, both initial and final states contain large prolate deformations on the order of $\beta_2 \sim 0.2$. When moving to the deformed Woods-Saxon potential, one expects large contributions from different orbitals, as all higher-spin orbitals contain $\Omega = 1/2$ projections. This is shown in Fig. \ref{fig:wf_comp_31S}, where the wave function composition of the single-particle state is shown graphically for the spherical and deformed Woods-Saxon potentials. Despite large mixtures in the wave function, the result remains practically unchanged with $b/Ac_1 = 5.18$. This is a feature generally seen for prolate deformed mirror nuclei, and is related to the relative sign of the $C_{j\Omega}$ components. An abbreviated version of the output in shown in Listing \ref{list:31S_output_nme}.

\begin{lstlisting}[caption={Excerpt from the output when running the \texttt{NME} package when calculating the weak magnetism contribution of the $^{31}$S mirror decay.}, label={list:31S_output_nme}, frame=single]
Nuclear potential: DWS
Transition from 31S [1/2] (0 keV) to 31P [1/2] (0 keV)
Neutron State
====================
Sorted single particle states
------------------------------
_____   1/2  -37.982427 MeV
....
_____   1/2  -13.317853 MeV<-------
....

Explicit wave function composition
------------------------------
Selected state: 1/2 (-13.3179 MeV)
Deformed oscillator quantum numbers: [211]
Orbital	C
1s1/2	-0.0175096
2s1/2	0.678097
....
1d3/2	0.570344
1d5/2	0.450627
....

Weak magnetism final result: b/Ac = 5.17988
\end{lstlisting}

\paragraph{$^{33}$Cl}

\begin{figure}[!ht]
\centering
\includegraphics[width=0.48\textwidth]{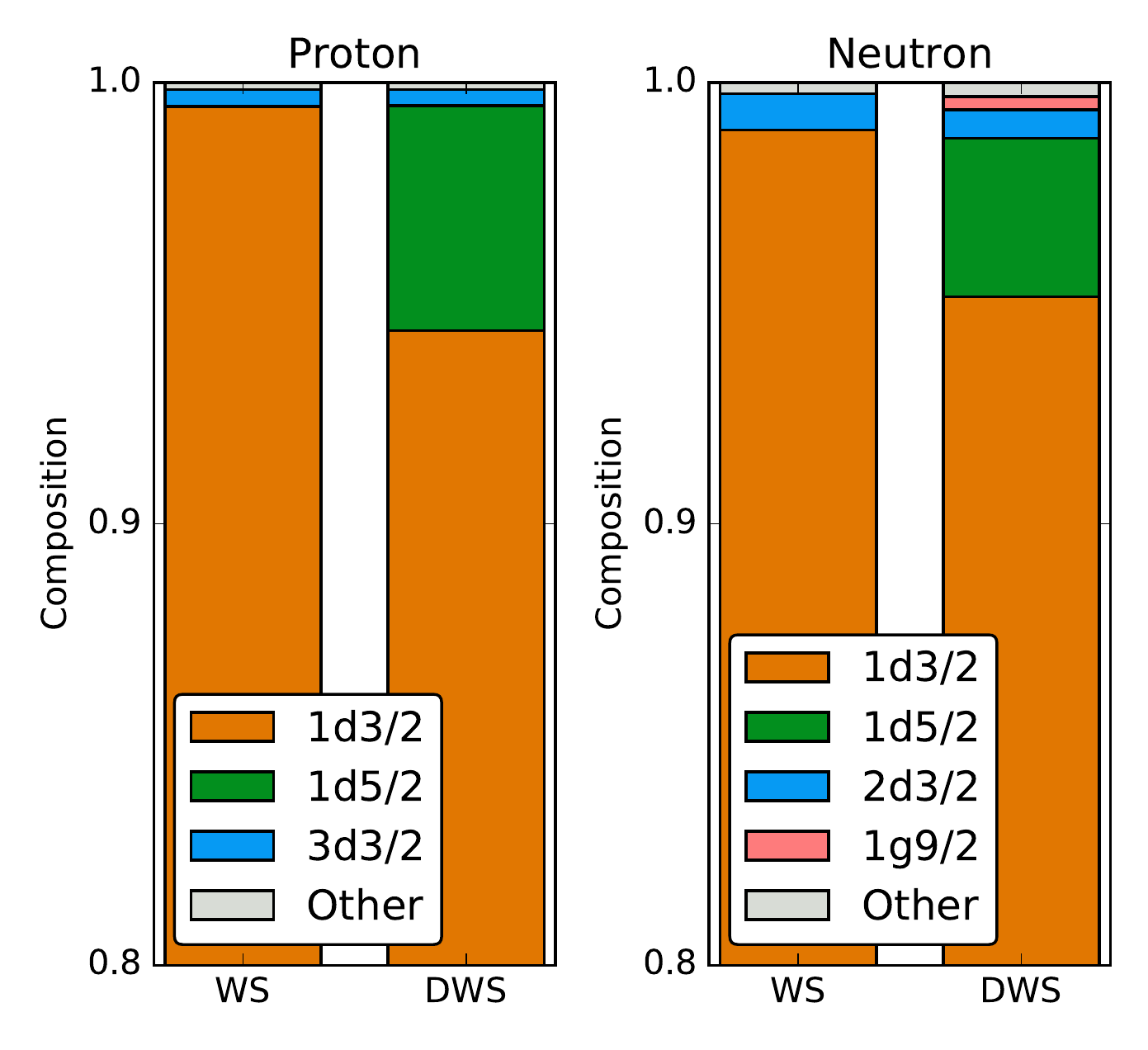}
\caption{Single particle wave function composition for the spherical Woods-Saxon (WS) and deformed Woods-Saxon (DWS) potentials for both proton (initial) and neutron (final) states of the $^{33}$Cl $\beta^+$ mirror decay. Admixtures of higher-lying orbitals in the deformed case are much smaller than those in $^{31}$S of Fig. \ref{fig:wf_comp_31S}, yet the value of $b/Ac$ is changed dramatically from $+2.46$ to $-1.13$ when moving from a spherical to a deformed Woods-Saxon potential. The reason for this lies in the relative signs of the different wave function components combined with the angular momentum coupling rules. This is frequently observed for oblate deformations.}
\label{fig:wf_comp_33Cl}
\end{figure}

Perhaps a more interesting example would be the $\beta^+$ mirror decay of $^{33}$Cl, only two nucleons away from our previous example. Using available experimental data one finds, however, a $b/Ac_1$ value of $-0.4589(55)$. This is unlike nearly all other mirror decays where values range between 2 and 7. Here the spherical harmonic oscillator and Wood-Saxon potentials fail spectacularly, both giving the same $b/Ac_1 = +2.46$ result. According to mean field results \cite{Moller2016}, however, both $^{33}$Cl and its daughter state $^{33}$S are rare cases of oblate deformation with $\beta_2 \sim -0.2$. Figure \ref{fig:wf_comp_33Cl} shows again the wave function composition in spherical and deformed Woods-Saxon potentials. While the admixtures of neighbouring orbitals are a lot smaller compared to the case of $^{31}$S - the scale now spans only 20\% - the new $b/Ac$ value is found to be $-1.13$, a profound result. For this to occur, the orbital matrix element, $M_L$, must exceed the regular Gamow-Teller matrix element, $M_{GT}$, by nearly a factor $-6$. This is an enhancement of a factor three compared to the analytical result of a pure $d_{3/2}$ orbital: $-l=-2$. 

Both examples serve to show that it is not necessarily the amplitude of the admixtures, $C_{j\Omega}^2$, but the relative sign and angular momentum couplings of the different contributions that matter.

\begin{figure}[!ht]
\centering
\includegraphics[width=0.8\textwidth]{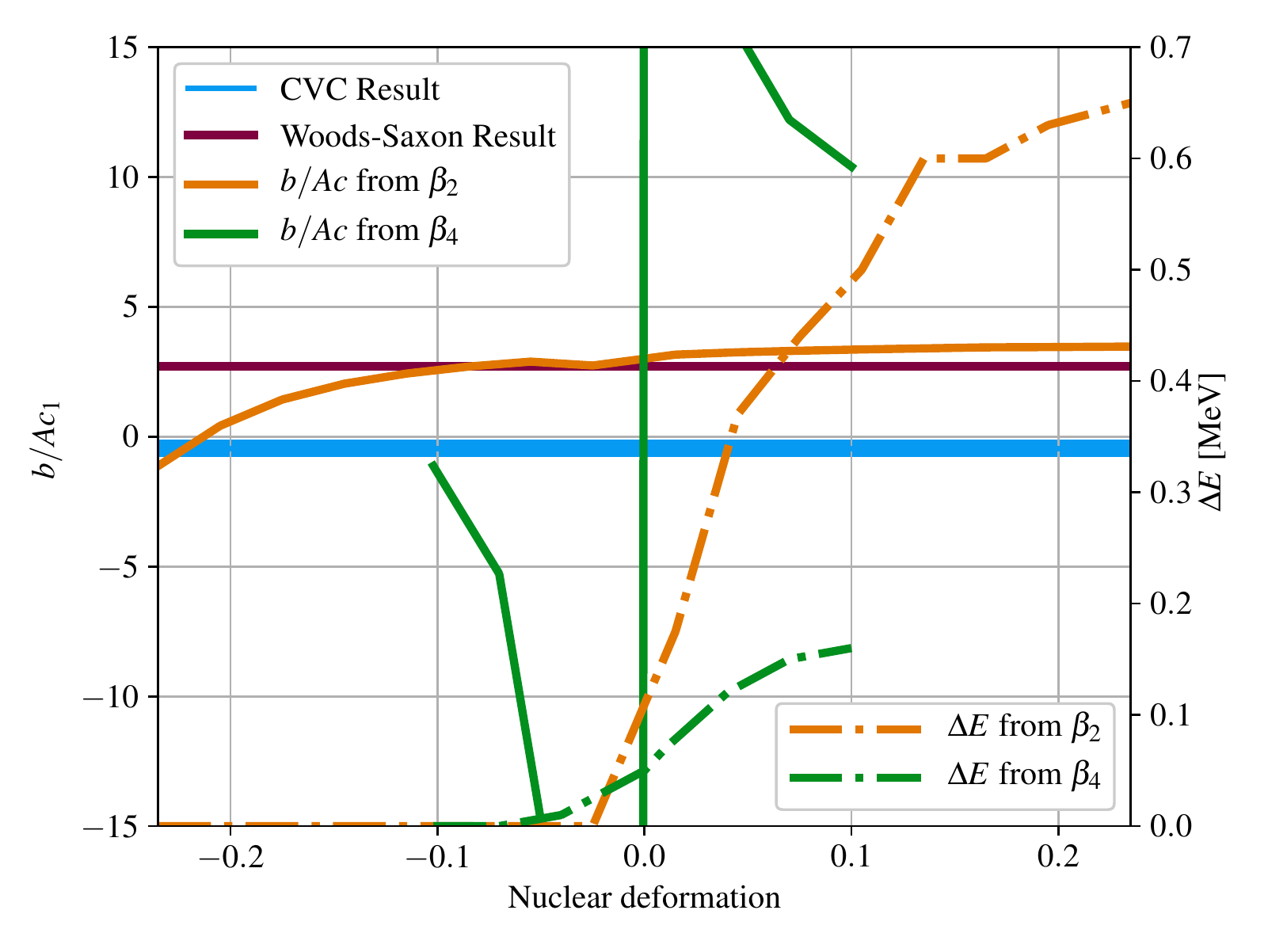}
\caption{Solid curves show the evolution of $b/Ac$ when changing either $\beta_2$ (orange) or $\beta_4$ (green) from its original value to that of opposite sign while dotted curves show the energy difference in MeV between the proposed single-particle state based on a filling of the calculated orbitals and the state corresponding to the right initial spin-parity. Initial values for $\beta_2$ and $\beta_4$ are -0.235 and -0.102, respectively. Both the experimental and spherical Woods-Saxon value are shown for reference. While the calculated $b/Ac$ value corresponds nicely to its experimental value (blue, uncertainty $\times 50$), differences in both $\beta_2$ and $\beta_4$ drastically change its value in oblate territory. As $\beta_4$ approaches zero, for example, a cancellation in the Gamow-Teller matrix element causes the ratio to quickly diverge. When using the mean-field results, the proposed single-particle state in the valence orbital carries the correct spin-parity, whereas discrepancies become larger as one move farther away from the mean-field results.}
\label{fig:bAc_vs_deformation}
\end{figure}

Given the somewhat surprising result obtained in $^{33}$Cl, it is worthwhile to look at the sensitivity of the calculation to the input deformations. Figure \ref{fig:bAc_vs_deformation} shows the value of the calculated $b/Ac$ value when changing $\beta_2$ or $\beta_4$ from their initial values to those of opposite sign while keeping all others constant. The vertical axis on the right shows the observed energy offset in MeV from the proposed single-particle state based on filling the calculated orbitals to that with the correct spin-parity. For reference, both the experimentally calculated and spherical Woods-Saxon value for $b/Ac$ are indicated.
Nuclear deformation is typically dominated by a quadrupole term, $\beta_2$. The slope in $b/Ac$ is largest at the value taken from mean-field calculations whereas it is nearly zero for its corresponding prolate deformation. This indicates a venture into oblately deformed territory can be hazardous as small differences in $\beta_2$ can result in drastic changes in the matrix element. When changing $\beta_4$, on the other hand, values for $b/Ac$ diverge quickly as $|\beta_4| \to 0$ for this particular case due to a cancellation in the Gamow-Teller matrix element.  Additionally interesting to note is that the correct single-particle state is automatically found for the mean field deformation values, while this is not any more the case once either $\beta_n$ diverges too much.

\subsubsection{Allowed $\beta$ decay matrix elements}
We move on to the calculation of the other nuclear form factors entering the $\beta$ spectrum shape. Three examples are provided in Table \ref{table:matrix_elements_calc}, where the relevant quantities for the allowed beta spectrum shape are written for the $^{35}$Ar mirror decay, $^{30}$S isospin triplet decay and simple $\beta$ decay candidate $^{106}$Ru. Based on the calculations from Ref. \cite{Moller2016}, the first has a clear oblate shape with a non-negligible hexadecopole moment of $\beta_4 = -0.113$. The $b/Ac_1$ value can be calculated to give $-0.8597(76)$ \cite{SeverijnsTBP} using experimental data, just like $^{33}$Cl far away from typical values between 2 and 7. Once again, the deformed Woods-Saxon potential delivers excellent results. Care must be taken, however, as also here results depend strongly on the deformation values.

\begin{table}[!ht]
\centering
\caption{Overview of the relevant nuclear matrix elements occurring in the allowed $\beta$ decay formulae. Extreme single-particle quantities are calculated in the three different possibilities for the nuclear potential: spherical harmonic oscillator (SHO), spherical Woods-Saxon (WS) and deformed Woods-Saxon (DWS). Nuclear shell model results were obtained using NuShellX@MSU \cite{Brown2014}. For $^{35}$Ar and $^{30}$S the full $sd$ shell was used with the USDB interaction \cite{Brown2006}. In the case of $^{106}$Ru, we used a truncated \texttt{jj45pn} model space. Due to the nature of the required coupling in allowed $\beta$ decay, only the $\pi g_{9/2}$, $\nu g_{7/2}, \nu d_{5/2}, \nu d_{3/2}$ and $\nu s_{1/2}$ were opened up and combined with the \texttt{jj45pna} interaction \cite{Hjorth-Jensen1995}. The only transition that contributes is $\nu g_{7/2} \to \pi g_{9/2}$, but inclusion of all positive neutron orbitals reduces the Gamow-Teller matrix element to the experimental value. Experimental $b/Ac_1$ values were taken from Ref. \cite{SeverijnsTBP}. For transitions within isospin multiplets, $d/Ac_1$ vanishes identically due to the Conserved Vector Current (CVC) hypothesis.}
{\renewcommand{\arraystretch}{1.3}
\begin{tabular}{c|l|lccc|c}
\hline \hline
Transition & Form factor ratio & SHO & WS & DWS & NSX@MSU & Reference\\
\hline
 & $b/Ac_1$ & 2.46 & 2.46 & -0.94 & 0.61 & -0.8597(76) \cite{SeverijnsTBP}\\
$^{35}$Ar $\to$ $^{35}$Cl & $d/Ac_1$ & 0 & -0.06 & -0.19 & 0 & 0 (CVC)\\
 & $\mathcal{M}_{121}^0/\mathcal{M}_{101}^0$ & -1.01 & -1.09 & -2.89 & -1.70 & -\\
\hline
 & $b/Ac_1$ & 5.18 & 5.18 & 5.25 & 5.99 & 6.11(33) \cite{SeverijnsTBP}\\
$^{30}_{16}$S $\to$ $^{30}_{15}$P & $d/Ac_1$ & 0 & 0.18 & -0.08 & 1.03 & 0 (CVC)\\
 & $\mathcal{M}_{121}^0/\mathcal{M}_{101}^0$ & 0 & 0 & 0.10 & -1.89 & -\\
\hline
 & $b/Ac_1$ & - & 4.21 & 5.58 & 4.21 & -\\
$^{106}_{44}$Ru $\to$ $^{106}_{45}$Rh & $d/Ac_1$ & - & -4.82 & -4.22 & -4.50 & -\\
 & $\mathcal{M}_{121}^0/\mathcal{M}_{101}^0$ & - & 0.35 & 0.10 & 0.30 & -\\
\hline \hline
\end{tabular}
}
\label{table:matrix_elements_calc}
\end{table}

Differences between potentials are minor for the $\beta^+$ decay of $^{30}$S in the isospin triplet, even though deformations are significant. In all filling schemes the wave function is dominated by the $s_{1/2}$ component, accompanied by large, and roughly equal, $d_{3/2}$ and $d_{5/2}$ projections in the DWS case. As $j=l-1/2$ orbital couplings tend to lower the $b/Ac$ value, while the reverse happens for $j=l+1/2$, the final value remains largely unaffected. An interesting situation occurs for the $^{106}$Ru[$0^+$] $\to {}^{106}$Rh[$1^+$] decay, with both initial and final states showing strong prolate deformations \cite{Moller2016}. It demonstrates both the flaws of the simple filling scheme and the use of enforcing specific single-particle spin states. Without this enforcement, the final proton state is put into the 1/2$^-$[301] orbital while the initial neutron state resides in the 3/2$^-$[541] orbital, whereas in the regular Woods-Saxon these fall into $1g_{9/2}$ and $1g_{7/2}$, respectively, i.e. orbitals with opposite parity. This conundrum can be resolved by looking at the spin state of neighbouring isotopes. The initial neutron spin state can be deduced from $^{105}$Ru, carrying spin +3/2. The final proton state, on the other hand, can be found from $^{105}$Rh which carries spin +7/2. Setting these constraints in the transition file through the option \texttt{ForcedSPSpin}, both states are found within 2 MeV of the initial estimate and the 3/2$^+$[411] and 7/2$^+$[413] orbitals are designated for neutron and proton, respectively. While the latter is dominated by the $1g_{9/2}$ orbital, the former contains only 30\% of $1g_{7/2}$ and large admixtures of all nearby $g$ and $d$ orbitals. Even though the nuclei are prolate deformed, the high level density for increasing $A$ makes complex behaviour possible. The resulting weak magnetism value is increased by 33\% (see Table \ref{table:matrix_elements_calc}), a potentially non-trivial result. In the \texttt{jj45pn} model space, only transitions between the spin-orbit partners of the $g$ orbitals, resulting in roughly the same values as for the pure Woods-Saxon potential. For all of these results, $g_A$ has been set to unity. In the heavier mass ranges, however, Gamow-Teller strengths have been known to be heavily quenched compared to shell model results. Note that, as this nucleus is situated in the middle of the fission fragment region, this result illustrates the important consequences of nuclear deformation on the reactor antineutrino anomaly \cite{Mueller2011, Hayes2016}.

\subsection{BSG}
\label{sec:BSG_example}
The core functionality of the code described here is the calculation of allowed $\beta$ spectrum shapes. As an example we consider the $^{67}$Cu ground state to ground state transition, $^{67}$Cu[3/2$^-$] $\to {}^{67}$Zn[5/2$^-$] ($Q=561.7\,$keV) which was recently used for a $\beta$-asymmetry parameter measurement \cite{Soti2014}. Both states show reasonable oblate deformation. Nonetheless, both spin states are reproduced automatically using the deformed Woods-Saxon potential. The neutron is put into the 3/2$^-$[301] state while the proton is in the 5/2$^-$[303] state, both of which are dominated by the $1f_{5/2}$ orbital. An abbreviated section of the spectrum output file is reproduced in Listing \ref{list:67Cu_output} using the matrix element results from Ref. \cite{Soti2014}.

\begin{lstlisting}[caption={Excerpt from the output from running the \texttt{BSG} package using shell model results from Ref. \cite{Soti2014} for the $\beta$ decay of $^{67}$Cu.}, label={list:67Cu_output}, frame=single]
Transition from 67Cu [-3/2] (0 keV) to 67Zn [-5/2] (0 keV)
Q Value: 561.7 keV      Effective endpoint energy: 561.693 keV
Process: B-     Type: Gamow-Teller
Partial halflife: 1.11294e+06 s
Calculated log ft value: 6.24694
Mean energy: 183.122 keV

Matrix Element Summary
------------------------------
b/Ac (weak magnetism)              : 2.72211
d/Ac (induced tensor)              : -1.37881
AM121/AM101                        : -0.306603

Spectral corrections
------------------------------
Phase space              : true
Fermi function           : true
....

Spectrum calculated from 0 keV to 561.693 keV with step size 0.1 keV

W [m_ec2]       E [keV]         dN_e/dW         dN_v/dW
1.000000        0.000000        0.000000        0.000000
1.000196        0.100000        0.000000        0.000000
1.000391        0.200000        0.715832        0.000000
1.000587        0.300000        1.960272        0.000001
....
\end{lstlisting}

\begin{figure}[!ht]
\centering
\includegraphics[width=0.5\textwidth]{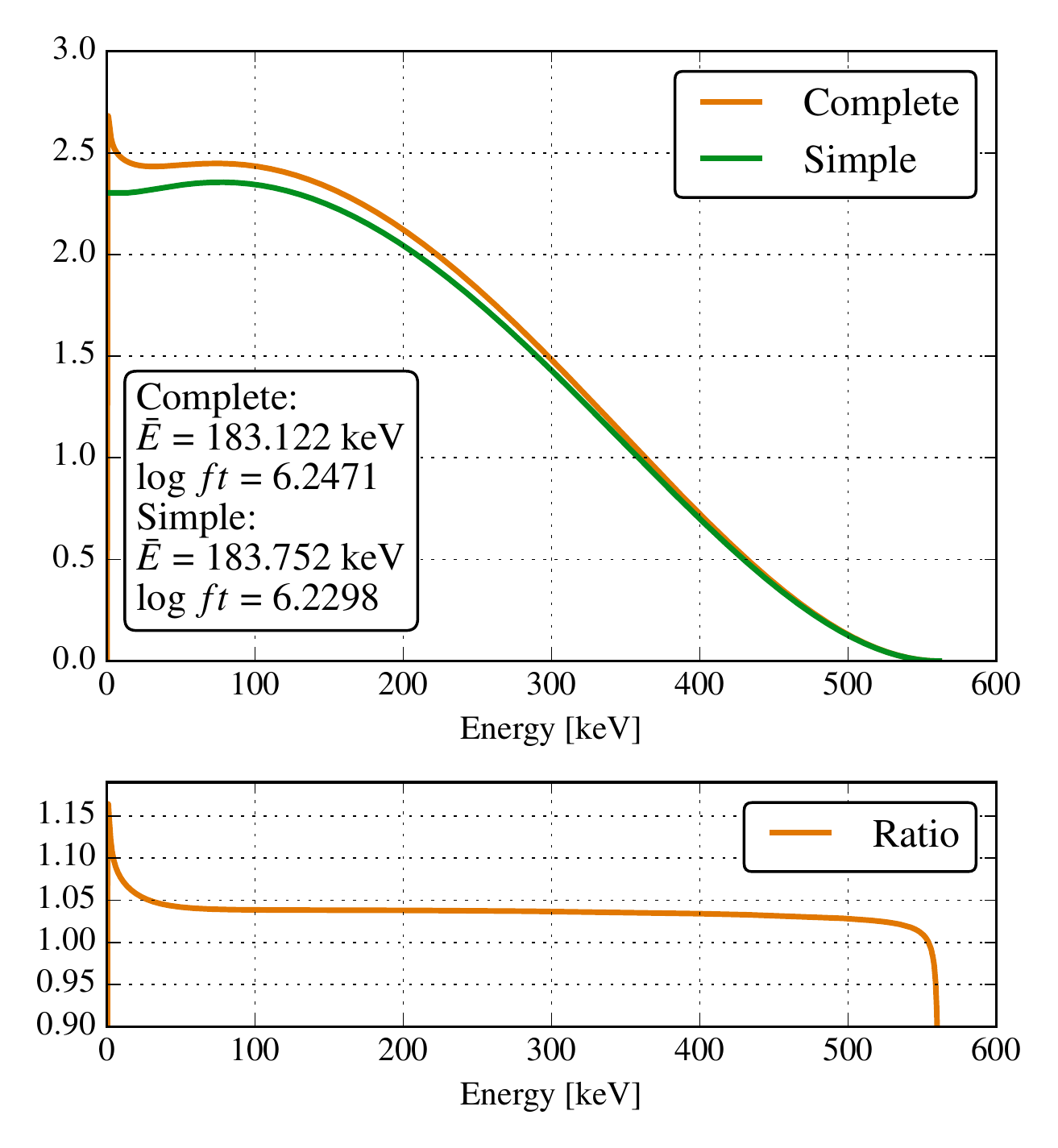}
\caption{Calculated $\beta^-$ spectra of the $^{67}$Cu $\to {}^{67}$Zn ground state to ground state transition, using only the phase space and Fermi function (Simple) and including all correction factors (Complete). The bottom panel shows the ratio between the latter and former over the full energy range. Due to the low endpoint energy, most of the ratio comes from constant factors. As a consequence, the mean energy is hardly changed in both calculations.}
\label{fig:spectrum_67Cu}
\end{figure}

The mean $\beta$ energy and $\log ft$ value are automatically included, and compared to externally provided data listed in the Evaluated Nuclear Structure Data File (ENSDF) \cite{ENSDF}. Both positron (electron) and (anti)neutrino spectra are given, taking into account differences in radiative corrections as discussed in Sec. \ref{sec:participating_corrections}. Figure \ref{fig:spectrum_67Cu} shows the spectrum including all corrections, and one using only the phase space and Fermi function factors. Due to the low endpoint energy it is the constant factors which have the largest contribution. As such, the ratio is nearly flat over the entire region, and differences in, e.g., the mean energy are small.

The usual suspects - finite size and radiative effects - are strongly dependent on $\alpha Z$ and $W_{(0)}$, with a significant increase in correction magnitude for increasing values of atomic number and energy. The atomic corrections such as exchange and screening, however, depend on the structure of the surrounding electron cloud and show non-trivial behaviour as a function of $Z$ \cite{Hayen2018}. The former in particular shows strong variations from one atom to the next, and displays clear atomic shell effects. As an example, we briefly consider the exchange magnitudes in the heavy Pb+ region. Exchange corrections rise to 30\% for the $^{241}$Pu $\to$ $^{241}$Am transition\footnote{This decay and the influence of exchange was for example discussed by Mougeot \emph{et al.} \cite{Mougeot2014}.}, while for $^{199}$Pt $\to$ $^{199}$Au decay (where the mother has a closed $5s$ atomic shell) the correction barely reaches 2\%. This has been extensively discussed in Ref. \cite{Hayen2018}, and is implemented as such in the code.

\begin{figure}[!ht]
\centering
\includegraphics[width=\textwidth]{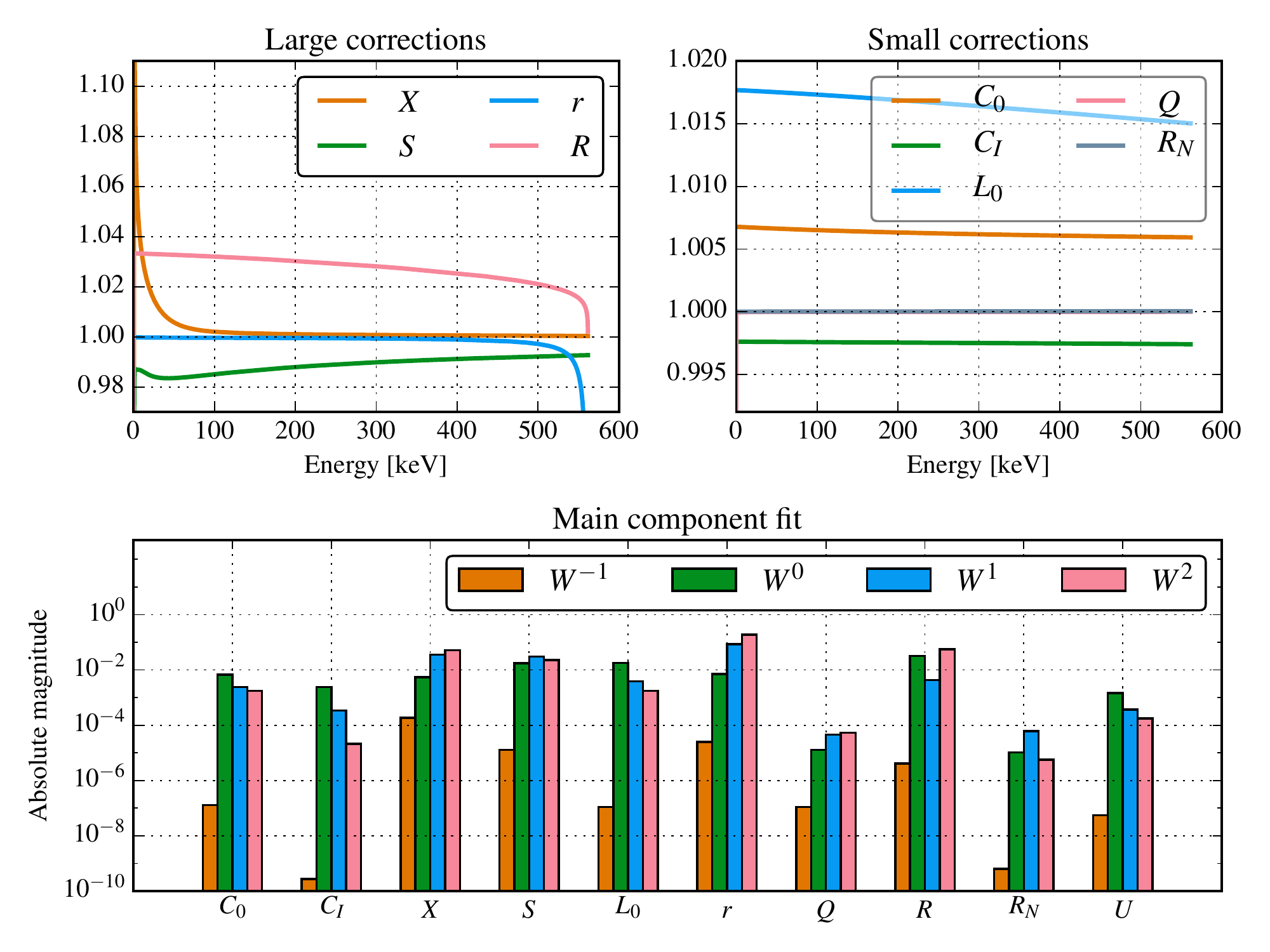}
\caption{Compilation of the correction factors used in the complete calculation of the $^{67}$Cu $\to {}^{67}$Zn ground state to ground state $\beta^-$ transition spectrum. An artificial split was created for the sole purpose of plotting in the specific case at hand. The bottom panel shows the result of a fit of all individual corrections using a simple $F(W) = 1 + a_0 + a_{-1}/W+a_1W+a_2W^2$ function, thereby showing the kinematical influences of all involved corrections. Notice the large logarithmic scale on the vertical axis.}
\label{fig:spectral_corrections_67Cu}
\end{figure}

To better understand the influence of the different correction factors, Fig. \ref{fig:spectral_corrections_67Cu} shows a summary of all terms. All correction factors were individually fitted using a simple $F(W) = 1 + a_0 + a_{-1}/W+a_1W+a_2W^2$ function, with $W$ the total electron energy in MeV, the results of which are shown in the bottom panel.

With this information it is more straightforward to extrapolate typical behaviour to higher energies and show the important corrections for different types of experiments. This is, however, only a rough approximation, as sudden onsets of deviation in correction factors are hard to capture using such a simple fit function.

\section{Applications}
\label{sec:applications}
As the experimental precision on $\beta$ decay variables steadily improves, the need for a high-fidelity theoretical calculation becomes all the more important. As such, the theoretical results presented here form the input for several different experimental campaigns in a variety of isotopes \cite{Perkowski2018, Broussard2019}. Beyond direct spectral measurements, the integrated form of the $\beta$ spectrum shape is a direct ingredient in the extraction of the $V_{ud}$ matrix element in both superallowed and mirror systems. The latter in particular is sensitive to nuclear structure information through induced currents. Currently the most accurate results are calculated numerically using nuclear shell model inputs \cite{Severijns2008, Towner2015}. The validity of the latter are, however, not guaranteed at the required level of precision for the next generation. Here this work can push forward the precision by allowing the user to easily change the values for the form factors of Eqs. (\ref{eq:bAc_BB})-(\ref{eq:Lambda_BB}) and replace them instead with experimentally measured values. This way, the resulting theoretical uncertainty in $f_A/f_V$ can be coupled to experimental data \cite{SeverijnsTBP}.

Additionally, the integrated spectrum shape value is an essential ingredient in astrophysical research such as the $r$-process \cite{Klapdor1985, Suzuki2012, Fang2013}. Here, typically only the Fermi function is included when calculating the Gamow-Teller strength function. Already from a transition reasonably close to stability such as the one discussed in Sec. \ref{sec:BSG_example}, differences of more than 10\% are observed in the $ft$ values. Despite significant nuclear structure uncertainties far away from stability, a systematic overestimation of the integrated phase space can have significant consequences, particularly for high masses.

Finally, a significant amount of interest is currently devoted to the so-called reactor antineutrino anomaly \cite{Mueller2011}. An accurate calculation of a tremendous amount of $\beta$ spectrum shapes lies at the heart of the analysis, and many approximations are introduced in the process \cite{Hayes2016}. Several of these, however, are not warranted for many of the transitions, leading to significant discrepancies \cite{Hayen2018b}. Several of the corrections reported on in this work have a significant effect on the final result of the reported anomaly, and are to be treated with extreme care.

\section{Conclusion}
\label{sec:conclusion}
Based on the allowed $\beta$ decay formalism recently described \cite{Hayen2018}, we developed two interconnected C++ libraries to perform both nuclear matrix element and spectrum shape calculations. With user-friendliness in mind, the calculations offer a multitude of options. A possibility for wave function input from many-body calculations was implemented to allow for nuclear matrix element calculations of the highest fidelity. Additionally, a graphical user interface was written in Python to further facilitate user interaction, and streamline connections to various databases. 

As several experimental groups are using allowed $\beta$ decay to further hone in on the specific structure of the weak interaction, the code presented here can be used as the theoretical input of the Standard Model to a few parts in $10^{4}$ precision. Additionally, it finds applications in the treatment of the reactor antineutrino anomaly, and can calculate $ft$ to a much higher precision than that typically provided by nuclear databases.

\section*{Acknowledgements}
The authors would like to thank L. De Keukeleere and S. Vanlangendonck for their valuable feedback. This work has been partly funded by the Belgian Federal Science Policy Office, under Contract No. IUAP EP/12-c and the Fund for Scientific Research Flanders (FWO).

\appendix

\section{Extreme single-particle evaluation}
In the simplest approach, initial and final nuclear states can be approximated by a single-particle configuration \cite{deShalit1974}. In this case the sum in Eq. (\ref{eq:operator_decomp}) contains only a single term, with $\alpha$ and $\beta$ single-particle states closest to the proton and neutron Fermi surfaces. The rest of the nucleus is considered to be an inert core with a possible rotation. The odd-$A$ nucleus wave function is trivial, while the state of even-$A$ nuclei is constructed using a spectator nucleon to obtain the correct angular momentum coupling.

In the absence of coupling to other, more advanced calculations, the program here described evaluates the nuclear matrix elements in the extreme single-particle approximation. As such, it is instructive to describe the ROBTD in this simplified situation. In the spherical case we can write down the non-trivial $C(L)$ angular momentum coupling factor for even-$A$ $\beta$ decays. For an even-even to odd-odd transition where the respective nuclear states are given by $|j_1j_1J_iM_iT_iT_{3i}\rangle$ and $|j_2j_1J_fM_fT_fT_{3f}\rangle$ the coupling is given by \cite{Rose1954, Behrens1982}
\begin{align}
C(L) &= \frac{\hat{J_i}\hat{J_f}\hat{T_i}\hat{T_f}}{\sqrt{1+\delta_{j_1j_2}}}(-1)^{T_f-T_{3f}}\left(
\begin{array}{ccc}
T_f & 1 & T_i \\
-T_{3f} & \pm 1 & T_{3i}
\end{array}
\right)
\left\{
\begin{array}{ccc}
\frac{1}{2} & T_f & \frac{1}{2}(T_f+T_i) \\
T_i & \frac{1}{2} & 1
\end{array}
\right\} \nonumber \\
& \times \sqrt{\frac{3}{2}} (-1)^L 2 [\delta_{j_1j_2}-(-1)^{j_1+j_2}] 
\left\{
\begin{array}{ccc}
j_2 & J_f & j_1 \\
J_i & j_1 & L
\end{array}
\right\}
\end{align}
where $\hat{j} = \sqrt{2j+1}$, $T$ stands for the total isospin and $T_{3}$ is the isospin projection on the 3-axis. An equivalent formula can be written down for odd-odd to even-even decays.

The situation in deformed nuclei is slightly different in that the individual angular momentum is not any more conserved, and is replaced with its projection along the symmetry axis. As such the single-particle matrix element undergoes some subtle changes, and the separation of the many-body nuclear matrix element into its many-particle coupling and single-particle matrix element is not always possible. We therefore discuss both odd-$A$ and even-$A$ nuclei in the extreme single-particle evaluation. Of primary concern is an understanding of the possible projection of the single-particle angular momentum along the symmetry axis. In the rotational ground state, one trivially has $K=\Omega$ for odd-$A$ nuclei. The matrix element for odd-$A$ decays is then written as \cite{Behrens1982}

\begin{align}
\langle &\phi(J_fK_f;\Omega_f) || O_{L} \tau^\pm || \phi(J_iK_i;\Omega_i) \rangle = \frac{\hat{J_i}\hat{J_f}}{\sqrt{(1+\delta_{K_f0})(1+\delta_{K_i0})}} \sum_{j_2j_1} C_{j_2\Omega_2}C_{j_1\Omega_1} \nonumber \\
&\times \left\{(-1)^{J_2-K_2+j_2-\Omega_2} \left( \begin{array}{ccc}
J_f & L & J_i \\
-K_f & \Omega_2-\Omega_1 & K_i
\end{array} \right) \left( \begin{array}{ccc}
j_2 & L & j_1 \\
-\Omega_2 & \Omega_2-\Omega_1 & \Omega_1
\end{array}\right)\right.  \label{eq:deformed_matrix_element_odd} \\
&\left. +  \left( \begin{array}{ccc}
J_f & L & J_i \\
K_f & -\Omega_2-\Omega_1 & K_i
\end{array}\right) \left(\begin{array}{ccc}
j_2 & K & j_1 \\
\Omega_2 & -\Omega_2-\Omega_1 & \Omega_1
\end{array} \right) \right\}\langle j_2 || O_{L} || j_1 \rangle. \nonumber
\end{align},
where $\Omega_i$ is the spin projection along the symmetry axis the single particle orbitals of spin $j_i$. Here we have already written the matrix element as it is implemented in this work by using the wave function expansion of Eq. (\ref{eq:wf_deformed_exp}). Equation (\ref{eq:deformed_matrix_element_odd}) gives rise to an additional selection rule as $L \geq |K_f-K_i|$ must hold, in addition to the usual angular momentum selection rule.

In the case of even-$A$ decays initial and final states are approximated by a coupling of two valence particles to the correct total angular momentum. In even-even nuclei the valence particles couple to $K=0$, while in odd-odd nuclei $K$ can be $|\Omega_p\pm\Omega_n|$. This sign ambiguity is discussed by Gallagher \emph{et al.} \cite{Gallagher1958} resulting in a set of coupling rules, implemented in this work. We have then \cite{Berthier1966}
\begin{align}
\langle &\phi(J_fK_f;\Omega_f) || \sum_{n=1,2} \{O_{L}\tau^\pm_n\} || \phi(J_iK_i=0;\Omega_i=0) \rangle = \frac{\hat{J_i}\hat{J_f}}{\sqrt{2(1+\delta_{K_f0})}} \left(\begin{array}{ccc}
J_f & L & J_i \\
-K_f & K_f & 0 
\end{array} \right)\nonumber \\
&\times [1+(-1)^{J_i}]\sum_{j_2j_1} C_{j_2-\Omega_2}C_{j_1\Omega_1} (-1)^{j_2-\Omega_2}  \left(\begin{array}{ccc}
j_2 & L & j_1 \\
-\Omega_2 & K_f & -\Omega_1
\end{array} \right)\langle j_2 || O_{L} || j_1 \rangle.
\label{eq:deformed_matrix_element_even}
\end{align}
Here $\langle j_2 || O_{L} || j_1 \rangle$ are simple spin-reduced single-particle matrix elements for the harmonic oscillator wave functions. 
%



\section*{References}
\bibliographystyle{elsarticle-num}
\bibliography{library}

\end{document}